\documentclass{Interspeech2024}

\interspeechcameraready

\usepackage{multirow}
\usepackage{multicol}
\usepackage{booktabs}
\usepackage{tabularray}
\usepackage{color}
\usepackage{xcolor}
\usepackage{array}
\usepackage{siunitx}

\newcommand{\et}{{et al.\,}}
\newcommand{\bds}{D3BV }

\title{DB3V: A Dialect Dominated Dataset of Bird Vocalisation \\for Cross-corpus Bird Species Recognition}

\name[affiliation={1, 3}]{Xin}{Jing}
\name[affiliation={2}]{Luyang}{Zhang}
\name[affiliation={2}]{Jiangjian}{Xie}
\name[affiliation={3}]{Alexander}{Gebhard}
\name[affiliation={5}]{Alice}{Baird}
\name[affiliation={1,3,4}]{Björn}{Schuller}
\address{
  $^1$Chair of Embedded Intelligence for Health Care \& Wellbeing, University of Augsburg, Germany
  $^2$School of Technology, Beijing Forestry University, China
  $^3$CHI -- Chair of Health Informatics, MRI, Technical University of Munich, Germany 
  $^4$Group on Language, Audio, \& Music, Imperial College London, UK
  $^5$ Hume AI, New York, USA}
\email{shyneforce@bjfu.edu.cn}

\begin{document}

\maketitle

\begin{abstract} 
% \blfootnote{* Corresponding author}
In ornithology, bird species are known to have variedit's widely acknowledged that bird species display diverse dialects in their calls across different regions. Consequently, computational methods to identify bird species onsolely through their calls face critsignificalnt challenges. 
There is growing interest in understanding the impact of species-specific dialects on the effectiveness of bird species recognition methods. Despite potential mitigation through the expansion of dialect datasets, the absence of publicly available testing data currently impedes robust benchmarking efforts. This paper presents the Dialect Dominated Dataset of Bird Vocalisation (\bds\kern-0.5em), the first cross-corpus dataset that focuses on dialects in bird vocalisations. The \bds comprises more than 25 hours of audio recordings from 10 bird species distributed across three distinct regions in the contiguous United States~(CONUS). In addition to presenting the dataset, we conduct analyses and establish baseline models for cross-corpus bird recognition. The data and code are publicly available online:
\ifinterspeechfinal
https://zenodo.org/records/11544734
\else
BLINDED
\fi

\end{abstract}

%BS: Do we have the image right for the map of the US picture?

\section{Introduction}
\label{sec:intro}
Bird monitoring is pivotal for biodiversity assessment and environmental conservation \cite{r2,r3}. Passive Acoustic Monitoring (PAM) technology, as outlined in prior studies \cite{r1}, has emerged as a key method for this purpose. In recent years, PAM devices have been extensively employed to capture bird vocalisations automatically. Subsequently, PAM technology facilitates species identification and diversity assessment through the analysis of these captured vocalisations~\cite{xie2023review}.

Long-term passive monitoring, however, will result in huge amounts of audio data, and with a scarcity in the skilled human resources required to effectively label this data automated workflows are of high importance~\cite{stowell2022computational}.
Although automatic bird recognition methods employing deep learning (DL) have become state-of-the-art for boosting data processing efficiency \cite{xie2023review}, bird recognition in the wild remains an open challenge.
 
The diversity observed in bird vocalisations underscores the existence of widespread dialects across different geographical regions \cite{aplin2019culture, smeele2024multilevel}. Research has identified microgeographical dialects, spanning less than 2\,km, which significantly impact the efficacy of automatic bird recognition methods \cite{r6, morgan2022open}.

These regional variations in vocalisations present a formidable challenge for improving the accuracy of recognition methods \cite{morgan2022open, r5, xie2023cross}, necessitating further investigation into cross-dialect recognition strategies.

The presence of dialects within a database parallels the challenge of dealing with disparate training and test sets stemming from distinct corpora or domains, thereby introducing a cross-corpus recognition hurdle. While expanding the collection of diverse dialect datasets may offer some relief, tackling the variability present in the testing set demands a more robust approach \cite{tan2024acoustic}. Numerous strategies have been explored to bolster the model's generalisation across various datasets \cite{conde2021weakly, li2023multimodal, zhang2023optimal, gebhard2023exploring, jing2022temporal}.

In scenarios like bird monitoring, where the specifics of the target domain are often uncertain, normalisation techniques can play a pivotal role in enhancing the model's generalisability. Instance normalisation (IN) \cite{Ulyanov2016Ins} stands as one such method, effectively mitigating instance-specific domain disparities and enhancing recognition performance across diverse tasks.

To compensate for the loss of classification information resulting from normalisation while simultaneously eliminating irrelevant device-specific nuances, approaches like audio residual normalisation \cite{kim2021domain} and relaxed instance frequency normalisation \cite{kim2022domain} have been employed, thereby bolstering the recognition model's generalisability across different audio sample devices. Tang \et \cite{tang2022acoustic} employed hybrid augmentation across diverse datasets to broaden the domain coverage of the training dataset, coupling it with per-channel energy normalisation (PCEN) to alleviate acoustic distortion and consequently improve cross-domain bird audio detection accuracy.
In a similar vein, Xie \et \cite{xie2023cross} devised an enhanced instance frequency normalisation strategy aimed at enhancing feature-specific expression, proving to be advantageous in cross-corpus recognition. These normalisation methodologies hold promise in mitigating dialect effects within bird vocalisation recognition.
 
\begin{figure}[ht]
    \centering
    \includegraphics[width=0.48\textwidth]{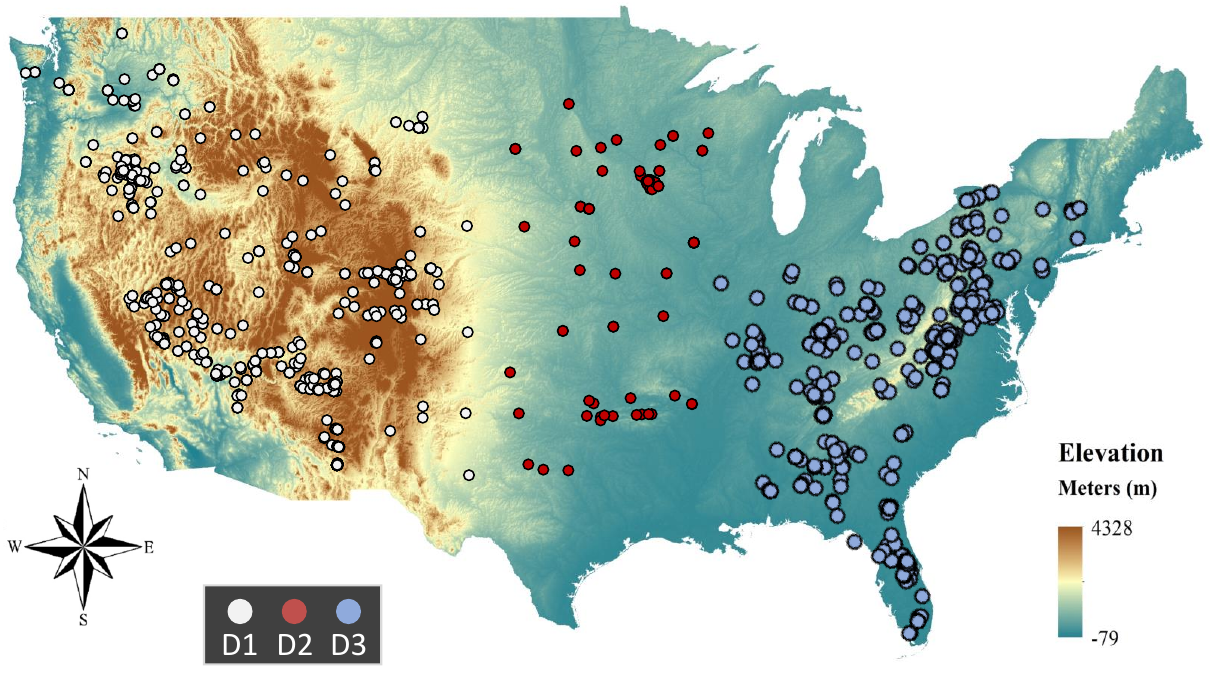}
    \caption{A map of the contiguous United States, illustrating three distinct geographical regions (D1-D3) delineated by variations in climate and vegetation patterns. Markers indicate co ordinates for the audio samples of \bds\kern-0.5em, with different marker colours representing different regions. The map was generated using ArcGIS.}
    \label{fig:region}
\end{figure}

However, before processing methodologies can be explored high-quality data remains fundamental to the development of any deep learning model.
% With this in mind, several datasets for audio-driven ornithological  machine learning tasks, including species recognition and general bird detection, have been developed, and we provide the most frequently employed in Table \ref{tab:existingdataset}.
% 
% \begin{table*}[t]
%     % \vspace*{-8mm}
%     \centering
%     \caption{Existing Bird vocalisation Datasets}
%     \label{tab:existingdataset}
%     \begin{tabular}{ll}
%        \hline
%        Tasks & Datasets\\
%        \hline
%        Bird species recognition  &  MLSP \cite{bib65}, ICML \cite{bib24}, NIPS4B \cite{bib26}, BirdCLEF \cite{bib66}, and Cornell Bird Challenge (CBC) \cite{bib67}\\
      % 
%        Bird individual recognition  & Little Owl, Murrelet, and Kiriwake's warbler \cite{bib72}, Great Cuckoo\cite{bib73}\\
       % 
%        Bird audio detetion  & BirdVox-DCASE-20k, Ff1010bird, and Warblrb10k\cite{stowell2018badchj}\\
       % 
%        Avian flight call detection  & Birdvox-full-night \cite{lostanlen2018birdvox}\\
       % 
%        Open set classification  & Soundscape of the Capital Region of New York State \cite{morgan2022open}\\
%        \hline
%     \end{tabular} 
% \end{table*}
% 
To the best of the authors knowledge, there is no bird vocalisation dataset that focuses on dialects\footnote{Datasets for audio-driven ornithological  machine learning tasks submitted as supplementary materia}. In this study, we establish a Dialect Dominated Dataset of Bird Vocalisations (\bds\kern-0.5em), which contains 3 sub-datasets from different geographic regions of the CONUS. 
The reason for this is that the CONUS has a different climatic distribution and a rather homogeneous biomes system, with Desert shrubs, Savanna, and Temperate broadleaved forests stretching from west to east (shown in Figure \ref{fig:region}). 
Considering that vegetation is a vital factor influencing bird activity, we separated the data into three distinct sections based on various plant types \cite{NAGDC}.
In addition, ten common bird species in the CONUS are selected to ensure their presence in all three regions.

%BS: xxx until here

We present a comprehensive overview of our dataset, and conduct a series of experiments aimed at discerning differences within bird vocalisations of identical species. Our experimental findings unveil the existence of dialects within the same bird species across different regions. Furthermore, the outcomes of our experiments, spanning several deep learning models, demonstrate the influence of regional dialects on model generalisation and robustness.

% url: https://www.reddit.com/r/MapPorn/comments/16fxc9b/map_showing_terrain_model_of_usa/

\begin{figure*}[t]
    \centering
    \vspace{-12mm}
    \includegraphics[width=0.9\textwidth]{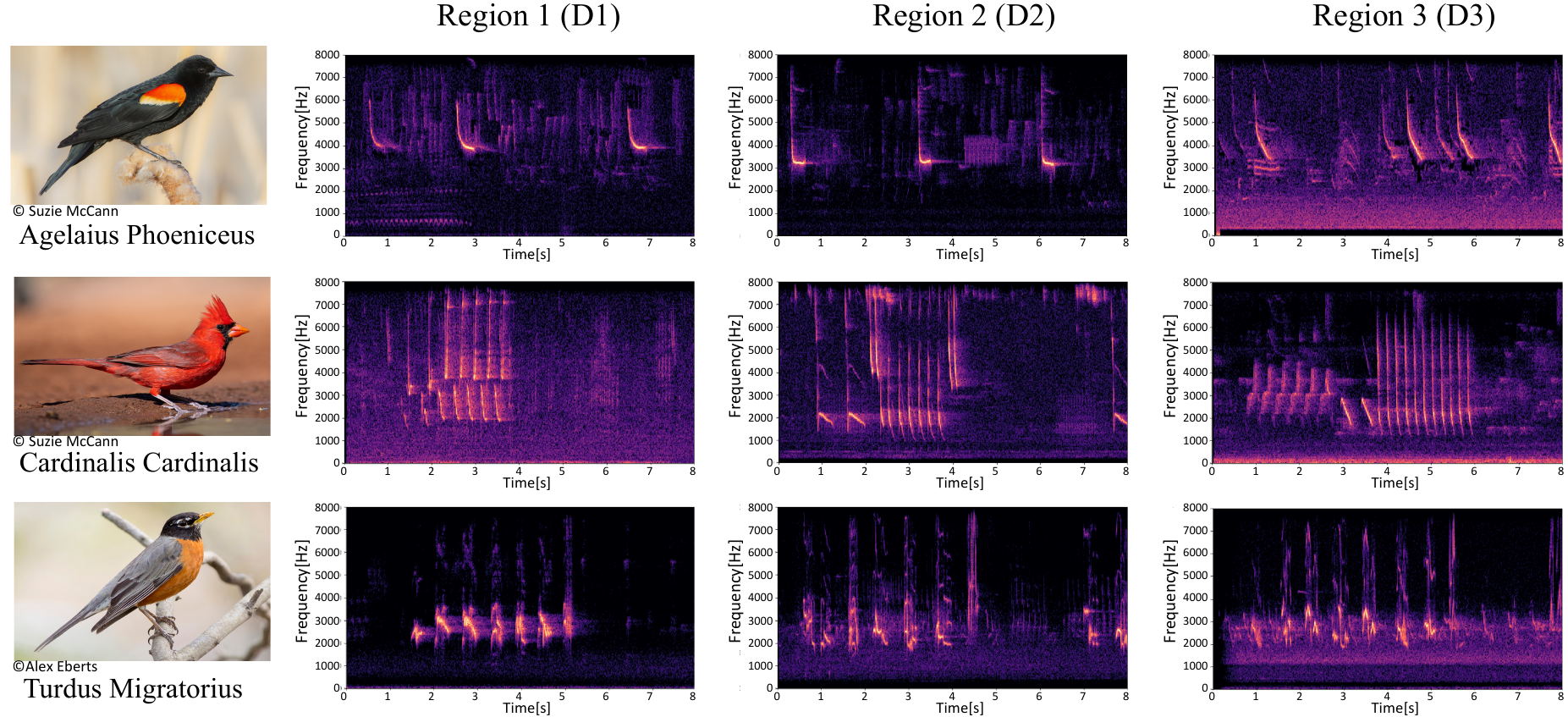}
    \caption{The spectrograms of three typical bird species in different regions across the contiguous United States in the \bds\kern-0.5em. The variations in frequency range, intervals, and other features of bird vocalisations indicate the existence of regional dialects within the same bird species.}
    \label{fig:dialects}
\end{figure*}

\begin{figure}[t!]
    \centering
    % \vspace{-8mm}
    \hspace*{-5mm}
    \includegraphics[width=0.48\textwidth]{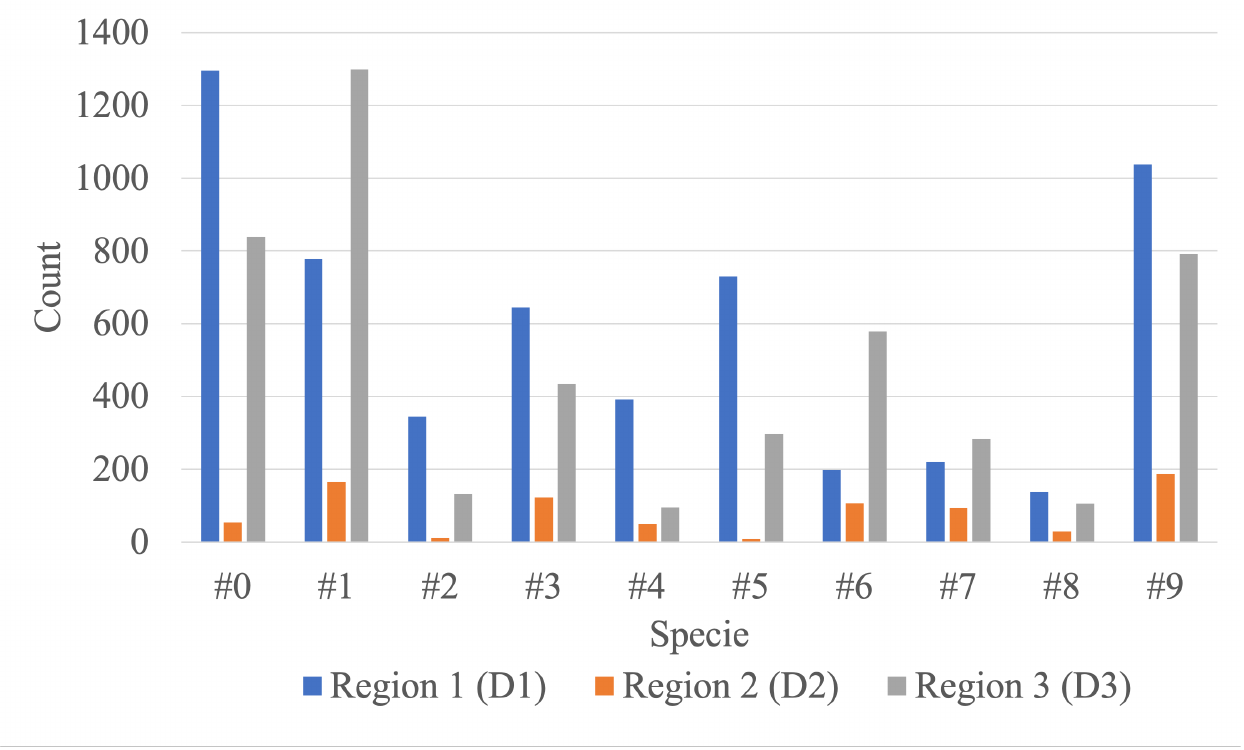}
    \caption{Amount of segments of different bird species across regions with Corresponding Codes (\#0\,--\,\#9) in Table \ref{tab:bird-info}}
    \label{fig:class_count}
\end{figure}
Our main contributions can be summarised as follows:

\begin{itemize}

%BS: Is the collection still ongoing? I changed to "collected", assuming it is closed by now.
\item we collected a dialect-dominated, preprocessed wild bird vocalisation dataset, enriched with detailed annotations.
%BS: reworded - please check:
\item we provide baseline models to benchmark the bird species recognition based on the cross dialects vocalisation dataset.

 \end{itemize}

To the best of our knowledge, \bds is the first open bird vocalisation dataset aimed at dialect effect study. We want to contribute to the field of bird vocalisation recognition by releasing our data set open-source and free for research use. 

\section{\bds Dataset}
\label{sec:data}
\subsection{Data Collection}
The \bds dataset encompasses audio clips from 10 bird species in three regions of the CONUS, amounting to a total duration of 91\,752 seconds. The bird vocalisation audio was gathered from "\textsc{Xeno-canto}"\footnote{https://xeno-canto.org/} , a platform dedicated to sharing wildlife sound recordings globally. The shared recordings are licensed under $\copyright$\,Creative Commons Attribution-NonCommercial-ShareAlike 4.0. Our selection criteria for bird species are based on their extensive distribution and the availability of high-quality audio data (labeled 'A' for sound quality by recordists) across all three regions, as indicated in the associated metadata.

When sourcing the data we chose to focus on the CONUS~(the latitude range of 24°N to 60°N)~ as the target area for the following reasons:

1. A wide range of bird species are found at all latitudes and longitudes, making it possible to investigate the dialects of the same species in different geographic regions.

2. The obvious pattern of regional variations in climate and vegetation can be broadly divided into three main geographical regions:

\begin{itemize}
    \item Western Cordillera (Region D1) ($\approx$100°W-130°W): This region is characterised by mountain ranges, including the Rocky Mountains and the Coastal Ranges. It extends along the western part of the contiguous United States.
    \item Interior Plains (Region D2) ($\approx$90°W-100°W): The vast and flat area known as the Great Plains lies between the Western Cordillera and the Appalachian Mountains. It is a region of extensive grasslands.
    \item Eastern Highlands (Region D3) ($\approx$70°W-90°W): The Appalachian Mountains, running along the eastern part of the continent, form the Eastern Highlands. This region consists of forested mountains and valleys.
\end{itemize}
% Based on the geographical regions above, our primary focus lies within the latitude range of 24°N to 60°N as our study region. 
We collected data corresponding to the regions outlined in Table \ref{tab:region-info}. Additionally, in Table \ref{tab:bird-info} we present comprehensive details of the 10 selected bird species, accompanied by the corresponding sample number across the three distinct regions. 

Figure \ref{fig:region} provides a visual representation of the recording locations, employing different marker colors—white, red, and blue—to denote D1, D2, and D3, respectively. Furthermore, Figure \ref{fig:dialects} offers a detailed examination of spectrogram disparities among three representative bird species across the three regions, revealing the presence of dialectical variations within the selected species.
Variations observed among regions exhibit distinct characteristics across species. Specifically, in the context of \textit{Agelaius phoeniceus}, there are varying frequencies of repetition in vocalisations. In the case of \textit{Cardinalis Cardinalis}, differences arise in the overall call durations, and for \textit{Turdus Migratorius}, variations are noted in the presence or absence of significant high-frequency harmonics.

\begin{table}
\centering
\caption{Distribution of audio clips across the regions.}
\label{tab:region-info}
\begin{tabular}{c | r | r } 
\toprule
Region & \# samples & \multicolumn{1}{c}{Longitude ranges}\\ 
\midrule
D1 & 5781 & (100.2167°W, 129.9241°W)\\ 

D2 & 831 & (90.3694°W, \,\,\,99.9995°W)\\ 

D3 & 4857 & (62.2240°W, \,\,\,89.9656°W) \\
\bottomrule
\end{tabular}
\end{table}

\begin{table*}[ht]
    \caption{Overview of the \bds dataset in terms of duration (seconds), frequency range (kHz), and number of audio files per species.}
    \label{tab:bird-info}
    \centering
    \hspace*{-8mm}
    \begin{tabular}{l|l|l|r|r|r}
        \hline
        \textbf{\rule{0pt}{2.5ex}\# Bird Species} & \textbf{Common name} & \textbf{Sound Type} & \textbf{Total Time (sec.)} & \textbf{Number of Segments} & \textbf{Range (kHz)} \\
        \hline
        \rule{0pt}{2.0ex}0 \textit{Agelaius phoeniceus} & Red-winged Blackbird & Songs & 17\,504 & 2188 & 2.8-5.7 \\
        1 \textit{Cardinalis cardinalis} & Northern Cardinal & Songs & 17\,944 & 2243 & 3.5-4.0 \\
        2 \textit{Certhia americana} & Brown Creeper & Calls & 3912 & 489 & 3.7-8.0 \\
        3 \textit{Corvus brachyrhynchos} & American Crow & Calls & 9624 & 1203 & 0.5-1.8\\
        4 \textit{Molothrus ater} & Brown-headed Cowbird & Calls & 4304 & 538 & 0.5-12.0 \\
        5 \textit{Setophaga aestiva} & American Yellow Warbler & Songs & 8288 & 1036 & 3.0-8.0 \\
        6 \textit{Setophaga ruticilla} & American Redstart & Songs & 7080 & 885 & 3.0-8.0 \\
        7 \textit{Spinus tristis} & American Goldfinch & Songs & 4784 & 598 & 1.6-6.7 \\
        8 \textit{Tringa semipalmata }& Willet & Calls & 2184 & 273 & 1.5-2.5 \\
        9 \textit{Turdus migratorius} & American Robin & Songs & 16\,128 & 2016 & 1.8-3.7  \\
        \hline
        \textbf{Total} & -& -& 91\,752 & 11\,469 & - \\
        \hline
    \end{tabular}
\end{table*}

\begin{table*}[t]
    \vspace*{-12mm}
    \caption{Cross-corpus baseline results, from four modeling strategies (see Section \ref{sec:experiment}. The evaluation is conducted by cross-corpus training and testing sets from the three distinct regions (D1-D3) and assessing their performance. We report accuracy (ACC), Unweight Average Recall (UAR), and F1-score (F1). \textbf{Bold} indicates the best performance on the cross-region vocalisation recognition}
    \label{tab:experimental-plan}
    \centering
    \begin{tabular}{cc ccc  cccc  cccc}
        \toprule
        \multirow{3}{*}{\textbf{Model}} & \multirow{3}{*}{\textbf{Train Dataset}} & \multicolumn{11}{c}{\textbf{Test Dataset}} \\ \cmidrule(lr){3-13}
        & & \multicolumn{3}{c}{\textbf{D1}} & \multicolumn{4}{c}{\textbf{D2}} & \multicolumn{4}{c}{\textbf{D3}} \\ \midrule
        % metrics
        & & \textbf{ACC} & \textbf{UAR} & \textbf{F1} 
        & &\textbf{ACC} & \textbf{UAR} & \textbf{F1} 
        & &\textbf{ACC} & \textbf{UAR} & \textbf{F1} \\ \midrule        
        %model 2
         \multirow{3}{*}{\textbf{TDNN+TN}} 
         & \textbf{D1}& 0.9707 & 0.9537 & 0.9664 && 0.5012 & 0.4018 & 0.3624 && 0.4633 & 0.4296 & 0.3939\\
         & \textbf{D2}& 0.2275 & 0.2308 & 0.1752 && 0.9762 & 0.9895 & 0.9889 && 0.3799 & 0.2545 & 0.2480\\
         & \textbf{D3}& 0.4978 & 0.4398 & 0.4226 && 0.6486 & 0.5397 & 0.4855 && 0.9631 & 0.946 & 0.9493\\ 
        \midrule
         %model 3
         \multirow{3}{*}{\textbf{TDNN+BN}} 
         & \textbf{D1}& 0.9621 & 0.9556 & 0.9550 && 0.6027 & 0.4907 & 0.4295 && 0.5768 & 0.5206 & 0.4900\\
         & \textbf{D2}& 0.2595 & 0.2864 & 0.2127 && 0.9048 & 0.8724 & 0.8801 && 0.4500 & 0.3367 & 0.3315\\
         & \textbf{D3}& 0.5704 & 0.5125 & 0.4930 && 0.7500 & 0.6209 & 0.5958 && 0.9693 & 0.9544 & 0.9479\\ 
        \midrule
        %model 4
         \multirow{3}{*}{\textbf{TDNN+GW}} 
         & \textbf{D1}& 0.9552 & 0.9518 & 0.9512 && 0.5906 & 0.4683 & 0.4006 && 0.5949 & 0.5366 & 0.5097\\
         & \textbf{D2}& 0.3045 & 0.3078 & 0.2269 && 0.9167 & 0.8509 & 0.8780 && 0.4885 & 0.3580 & 0.3357\\
         & \textbf{D3}& 0.5948 & 0.5363 & 0.5126 && 0.7488 & \textbf{0.6294} & 0.5917 && 0.9570 & 0.9287 & 0.9369\\ 
        \midrule
        %model 5
         \multirow{3}{*}{\textbf{TDNN+IFN}} 
         & \textbf{D1}& 0.9052 & 0.8721 & 0.8734 &  & \textbf{0.7246} & \textbf{0.6262} &\textbf{ 0.5837} &  & \textbf{0.6979} & \textbf{0.6313} & \textbf{0.6073}\\
         & \textbf{D2}& \textbf{0.3628} & \textbf{0.3651} & \textbf{0.2641} && 0.8214 & 0.6664 & 0.6565 && \textbf{0.5086} & \textbf{0.4085} & \textbf{0.3521}\\
         & \textbf{D3}& \textbf{0.6439} &\textbf{ 0.5740} & \textbf{0.5540} && \textbf{0.7766} & 0.5838 & \textbf{0.5918} && 0.8852 & 0.8217 & 0.8489\\ 
        
        \bottomrule
    \end{tabular}
\end{table*}
\subsection{Data preprocessing}
All audio files are converted to the Waveform Audio File Format (WAV) with a mono channel. As depicted in Table \ref{tab:bird-info}, the majority of bird vocalisations exhibit a frequency range within [0-8\,kHz]. Although \textit{Molothrus ater} has a maximum frequency of 12\,kHz, our analysis reveals that all instances of \textit{Molothrus ater} in our dataset had their highest frequencies below 8\,kHz. Consequently, we uniformly set the sampling frequency to 16\,kHz.

%To address variable audio lengths and optimise feature extraction, w
We adopt a method inspired by \cite{morgan2022open}, clipping the audio into 8-second segments. For segments shorter than 8 seconds, zero-padding is applied.
The segments are then thoroughly manually checked to validate the presence of bird vocalisation lasting more than one second in each segment. This filtering strategy improves the dataset's robustness and reliability while also aligning with the suitable preprocessing procedures for subsequent analysis.
Figure \ref{fig:class_count} lists the number of segments of different bird species across regions.

\section{Baseline Experiment}
\label{sec:experiment}

\subsection{Baseline and Compared Model}

Prior research has demonstrated that Time-Delay Neural Networks (TDNNs) and their variations \cite{xie2023cross}, exhibit superior performance compared to alternative methods in the realm of cross-corpus recognition of bird vocalisations. Therefore, to offer a baseline for other to benchmark the \bds dataset against, we apply four TDNN-based models with different normalisation methods in this work.
\begin{itemize}
    \item \textbf{TDNN + TN}~\cite{xie2023cross}: Time-Norm normalisation, which normalises each log-Mel spectrogram on the temporal axis, is applied to replace the BN layer in the TDNN + BN model.
    
    \item \textbf{TDNN + BN}~\cite{xie2023cross}: Compared to the original TDNN~\cite{fan2021deep}, Batch Normalisation (BN) and a Rectified Linear Unit (ReLU) layer are appended after each Conv1D layer before the pooling layer.
    
    \item \textbf{TDNN + GW}~\cite{xie2023cross}: GW~\cite{Huang2021Group}  efficiently leverages whitening benefits, maintaining a balance between learning efficiency and representational capacity.
    
    \item \textbf{TDNN + IFN}~\cite{xie2023cross}: In \cite{xie2023cross}, it is demonstrated that the frequency of a vocalisation varies more significantly between different domains in the context of bird species recognition. Therefore, the Instance Frequency normalisation (IFN) was proposed to eliminate the instance-specific contrast information on the frequency dimension for a more efficient extraction of domain-invariant features. The formula of IFN is:
    \begin{equation}
        IFN(x)=\frac{x-\mu_{in}}{\sqrt{\sigma_{in} + \delta}}, 
    \end{equation}
    where, $\mu_{in}, \sigma_{in}\in \mathbb{R}^{NxF}$ are the mean and standard deviation of the input feature $x\in\mathbb{R}^{N\times F \times T}$, in which $N, F$, and $T$ are batch size, frequency dimension, and time dimension respectively. $\delta$ is a small number to improve numerical stability.
   
\end{itemize}

%BS: where is a validation set definition? I think generally, you could write 2 more sentences about the partitioning - for example when n=m - how is the set divided? Is it somehow stratified, e.g., by geographic aspects?

\subsection{Experiment Setups}
From each audio sample, we extract Log Mel spectrograms with an FFT size of 2048, a hop length of 512, and 128 Mel bins. 
As shown in Table \ref{tab:experimental-plan}, we conduct the experiments across regions by setting D$n$D$m$ ($n,m\in(1, 2, 3)$). When $n=m$~(in-region), training and testing occur on the same region data, while $n\neq m$~(cross-region) indicates different regions involved in the training and testing.
For in-region recognition, the sub-dataset is partitioned into an 8:1:1 ratio of training, validation, and test sets. When referring to cross-region recognition, the data within a single regional dataset will not be split like the in-region dataset setup. The bird data from the current region~(D$n$) will be utilised as training data, while all data from the other two regions~(D$m, n\neq m$) will serve as test data respectively.
The loss function is Cross Entropy function.
Training is conducted with a batch size 32 and the learning rate is 1e-4, for 40 epochs. The Adam optimiser with 0.8 decay every 3 steps is applied for all the models.

\subsection{Result and Analysis}
The results, as summarised in Table \ref{tab:experimental-plan}, consistently demonstrate that performances within the same region (\textit{D}$n$\textit{D}$n$) generally surpass those across different regions (\textit{D}$n$\textit{D}$m$) across all models. This observation strongly suggests the presence of dialectal variations among regions.

Furthermore, the performance disparities observed in models trained on \textit{D1D2} and \textit{D1D3} datasets vividly underscore the influence of geographical distance on bird dialects. Notably, as the distance from the training data collection area increases, there is a noticeable degradation in model performance. These findings align with previous research \cite{lewis2021uses}, which suggests that dialectal variations often correlate with the geographical separation of bird populations.

It's worth mentioning that models trained on \textit{D2} data exhibit the least effective performance when tested on cross-region datasets. This is primarily attributed to \textit{D2} being situated in a region characterized by grasslands and relatively lower bird population density than the other two regions, resulting in a smaller number of audio samples in the \textit{D2} dataset.

Moreover, comparing normalisation methods, it's evident that Batch Normalisation (BN) normalises samples within a single batch, thereby potentially enhancing the model's generalisation ability. Conversely, Group Whitening (GW) normalises samples within constructed groups, offering improvements over BN by addressing within-batch normalisation limitations, ultimately enhancing generalisation 

%when compared to BN.
Both BN and GW normalise the time and frequency dimensions simultaneously, whereas Temporal Normalisation (TN) and Instance Frequency Normalisation (IFN) independently normalise time and frequency dimensions. With this in mind, we observe that TN exhibited the poorest performance, while IFN demonstrated the highest. This suggesting that frequency normalisation plays a stronger role in mitigating the influence of dialect and enhances the cross-corpus recognition capability of models. These findings align with existing literature \cite{Stowell2014large}, which suggests that birds adapt to diverse acoustic environments through Frequency Modulation (FM), underscoring the significance of frequency variations as a fundamental characteristic of dialectal differences.
Nevertheless, the performance of IFN requires further improvement, and it is possible that taking the time dimension of variation into consideration can aid this. Drawing on literature \cite{kim2022domain}, relaxed instance frequency normalisation may be a potential approach to implement next.

\section{Conclusions}
\label{sec:conclusion}

%BS: conclusion is usually held in past tense - I changed.
This paper introduces D3BV, a novel dataset of bird vocalizations designed to investigate the influence of dialects on bird species recognition. Baseline models have been established to benchmark cross-corpus bird species recognition based on vocalizations. We anticipate that the D3BV dataset and our findings will serve as a foundation for future research in bird vocalization recognition. We advocate for a comprehensive examination of bird dialects, emphasizing an ornithological perspective, and believe this dataset marks the first step toward a thorough quantitative investigation of bird dialects.

\section{Acknowledgements}

\ifinterspeechfinal
     This work was funded by the China Scholarship Council (CSC), Grant \#\,202006290013, and by the DFG, Reinhart Koselleck-Project AUDI0NOMOUS (Grant No.\ 442218748), and the National Natural Science Foundation of China (No.\ 62303063). Thanks to Zhexiu Yu, a PhD student at the School of Forestry, Beijing Forestry University, for helping with the of the map of the CONUS.
\else
     The authors appreciate the valuable contributions of discussion participants and the support that facilitated the development of this research.
\fi
\bibliographystyle{IEEEtran}
\bibliography{mybib}

\end{document}

\section{Supplementary material -- Datasets}
This appendix documents the audio-driven ornithological  machine learning tasks and related datasets.
\begin{table}[h]
    \centering
    \caption{Existing Bird vocalization Datasets}
    \label{tab:existingdataset}
    \begin{tabular}{ll}
       \hline
       Tasks & Datasets\\
       \hline
       Bird species recognition  &  MLSP \cite{bib65}, ICML \cite{bib24}, NIPS4B \cite{bib26}, BirdCLEF \cite{bib66}, and Cornell Bird Challenge (CBC) \cite{bib67}\\
       
       Bird individual recognition  & Little Owl, Murrelet, and Kiriwake's warbler \cite{bib72}, Great Cuckoo\cite{bib73}\\
       
       Bird audio detetion  & BirdVox-DCASE-20k, Ff1010bird, and Warblrb10k\cite{stowell2018badchj}\\
       
       Avian flight call detection  & Birdvox-full-night \cite{lostanlen2018birdvox}\\
       
       Open set classification  & Soundscape of the Capital Region of New York State \cite{morgan2022open}\\
       \hline
    \end{tabular} 
\end{table}

\section{\refname}
\printbibliography[heading=none]